\documentclass[journal,twoside,web]{ieeecolor}
\usepackage{tmi}
\usepackage{cite}
\usepackage{amsmath,amssymb,amsfonts}
\usepackage{algorithmic}
\usepackage{color}
\usepackage{makecell}
\usepackage{graphicx}
\usepackage{textcomp}
\usepackage{diagbox}
\usepackage{amsmath}
\usepackage{autobreak}
\usepackage{bbding}
\usepackage{multirow}
\usepackage{hyperref}
 \usepackage{mathrsfs}
 \usepackage[ruled,vlined]{algorithm2e}
 
\definecolor{ref}{rgb}{0,0.541,0.855}
\def\BibTeX{{\rm B\kern-.05em{\sc i\kern-.025em b}\kern-.08em
    T\kern-.1667em\lower.7ex\hbox{E}\kern-.125emX}}
\begin{document}
\title{Zero-shot Medical Image Translation via Frequency-Guided Diffusion Models}
\author{Yunxiang Li, Hua-Chieh Shao, Xiao Liang, Liyuan Chen, Ruiqi Li, Steve Jiang, Jing Wang, You Zhang
\thanks{This work was supported by the National Institutes of Health (Grant No. R01 CA240808, R01 CA258987). (Corresponding author: You Zhang.) }
\thanks{Yunxiang Li, Hua-Chieh Shao, Xiao Liang, Liyuan Chen, Ruiqi Li, Steve Jiang, Jing Wang, You Zhang are with Medical Artificial Intelligence and Automation (MAIA) Laboratory, Department of Radiation Oncology, UT Southwestern Medical Center, Dallas, TX, 75390, USA (e-mail: yunxiang.li@utsouthwestern.edu, hua-chieh.shao@utsouthwestern.edu, xiao.liang@utsouthwestern.edu, liyuan.chen@utsouthwestern.edu, ruiqi.li@utsouthwestern.edu, steve.jiang@utsouthwestern.edu, jing.wang@utsouthwestern.edu, you.zhang@utsouthwestern.edu).}
}
\maketitle

\begin{abstract}
Recently, the diffusion model has emerged as a superior generative model that can produce high quality and realistic images. However, for medical image translation, the existing diffusion models are deficient in accurately retaining structural information since the structure details of source domain images are lost during the forward diffusion process and cannot be fully recovered through learned reverse diffusion, while the integrity of anatomical structures is extremely important in medical images. For instance, errors in image translation may distort, shift, or even remove structures and tumors, leading to incorrect diagnosis and inadequate treatments. Training and conditioning diffusion models using paired source and target images with matching anatomy can help. However, such paired data are very difficult and costly to obtain, and may also reduce the robustness of the developed model to out-of-distribution testing data. We propose a frequency-guided diffusion model (FGDM) that employs frequency-domain filters to guide the diffusion model for structure-preserving image translation. Based on its design, FGDM allows zero-shot learning, as it can be trained solely on the data from the target domain, and used directly for source-to-target domain translation without any exposure to the source-domain data during training. We evaluated it on three cone-beam CT (CBCT)-to-CT translation tasks for different anatomical sites, and a cross-institutional MR imaging translation task. FGDM outperformed the state-of-the-art methods (GAN-based, VAE-based, and diffusion-based) in metrics of Fréchet Inception Distance (FID), Peak Signal-to-Noise Ratio (PSNR), and Structural Similarity Index Measure (SSIM), showing its significant advantages in zero-shot medical image translation.

\end{abstract}


\section{Introduction}
\label{sec:introduction}
\IEEEPARstart{M}{edical} image-to-image translation aims to transfer images from one domain to another while preserving the structural integrity. One of the applications is CBCT-to-CT translation\cite{angelopoulos2012comparison}, which is critical for radiotherapy applications, including image de-noising, dose calculation and accumulation, and treatment plan adaptation. Currently, most image translation methods are based on the generative adversarial networks (GAN)\cite{goodfellow2020generative} or the variational autoencoders (VAE)\cite{kingma2019introduction}. CycleGAN\cite{zhu2017unpaired} is widely used in medical image translation tasks, which learns through two neural networks playing against each other, indirectly capturing information to provide an implicit representation about the distribution of both source and target domains. However, such implicit representations are prone to learning biases, including premature convergence of the discriminator and pattern collapse, which affect the quality of the synthesized images. Recently, diffusion models are developed as a new type of generative AI method, which can generate images of very high quality with excellent Frechet Inception Distance (FID)\cite{heusel2017gans,dhariwal2021diffusion,yang2022diffusion,song2021scorebased,song2021denoising}. One representative diffusion model, Denoising Diffusion Probabilistic Model (DDPM)\cite{ho2020denoising}, is trained to denoise samples corrupted by varying degrees of Gaussian noise. The noise-corrupted samples are generated by a Markov chain Monte Carlo (MCMC) process called forward diffusion, by gradually corrupting an image with increasing levels of white Gaussian noise. Then, based on the corrupted images, DDPM progressively denoises and transforms the noise into meaningful, high-quality images by learning to reverse the forward diffusion process (reverse diffusion), based on Langevin dynamics. The forward and reverse diffusion processes help the DDPM to learn the underlying latent space and distribution of the images, and to generate new images from the learned distribution.

Considering diffusion models can generate images that outperform GAN-based models in terms of realism and quality across the board, the interest in applying diffusion models for medical image translation is continuously growing. By DDPM, we could potentially perform the forward diffusion on the source image domain, and use a learned reverse diffusion to convert the noise-corrupted image into the target image domain, to achieve direct domain translation. Such a strategy has found success in natural image translation, such as cat-to-dog image translation, which only requires the preservation of some domain-independent features (pose, color, etc.), such as in \cite{meng2021sdedit,zhao2022egsde}. However, the use of diffusion models in medical image-to-image translation tasks is currently limited, as its forward diffusion process on the source image leads to structural detail losses, which cannot be fully recovered in the reverse denoising process. The geometrical and structural integrity of the translated (synthesized) images is especially vital for medical applications like surgical planning and radiotherapy, as medical images, rich in accurate anatomical details, are crucial for both procedures. They guide surgeons in mapping out procedures and help in planning precise radiation doses for cancer treatment. A compromise in anatomical accuracy could lead to serious outcomes, including potentially harming the patient or inadequately treating the disease.
To preserve the structural information, some works\cite{lyu2022conversion} use paired images to develop diffusion models by feeding source images as a condition into the reverse diffusion process to constrain the solution of target domain images. However, such an approach requires curated source and target image pairs with an matching anatomy, which are scarce and difficult to obtain in medical imaging due to inter-scan anatomical motion, imaging cost, radiation doses, and etc. Using the source image itself as the condition also introduces domain-specific information that may make the diffusion model less robust. For instance, if the anatomical sites changed for the source image, or a different imaging protocol has been used for the source image, the out-of-distribution source image may reduce the inference accuracy of the diffusion model during the testing stage.

For diffusion models, the forward diffusion process can be considered as a low-pass filter (see section III.A). In contrast, the anatomical structure outlines in medical images are mostly embedded as high-frequency information, which gets filtered through the forward diffusion process. For the same anatomy, if the differences between two image domains/batches are mostly at the intermediate frequencies in the Fourier domain, we can potentially extract high-frequency information from the source image domain as a structural prior to condition target domain image generations. We can use a diffusion model trained solely via the target domain images to fill the intermediate-frequency vacancy. For the CBCT-to-CT translation problem, an important task in radiotherapy, we quantitatively and statistically analyzed the difference between CBCT and CT images in the frequency domain and found that the main differences between them are at intermediate frequencies, shown in Fig. \ref{fig:frequency_analysis}. Built upon this frequency-domain observation, we designed an image translation diffusion model jointly guided by a high-pass filter and a low-pass filter (a controlled forward diffusion process), in which the low-pass filter captures intensity and overall semantic information, and the high-pass filter captures anatomical details. The resulting frequency-guided diffusion model (FGDM) is conditioned by high-frequency information and low-frequency information to generate the intermediate-frequency information for medical image translation. 

Compared to GANs or VAEs, FGDM offers an additional benefit as it only uses the data from the target domain to train the diffusion model to learn the distribution at intermediate frequencies. The resulting model can be used for image translation from a different source domain without any fine-tuning, if the source domain images share similar information at both frequencies ends with the target domain images (Fig. \ref{fig:frequency_analysis}). The loosened restrictions on the source domain allow natural zero-shot translation\cite{romera2015embarrassingly}, by which the model can be applied to source images in other domains or distributions without transfer learning or test-time re-training. 
\begin{figure}[ht]
\centering
  \includegraphics[width=.5\textwidth]{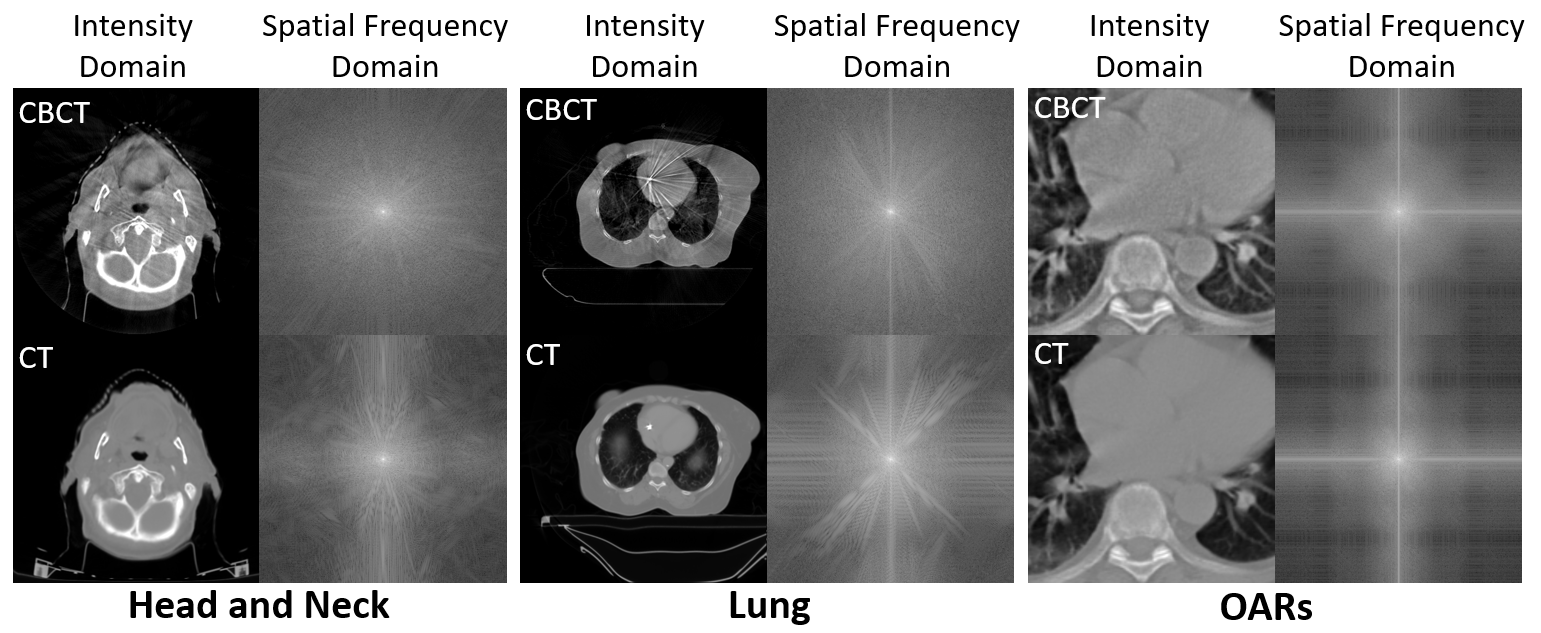}
  \caption{Examples of intensity domain and spatial frequency domain visualizations of three pairs of CBCT images and CT images from different datasets.}
  \label{fig:frequency_analysis}
\end{figure}

\begin{figure}[ht]
\centering
  \includegraphics[width=.5\textwidth]{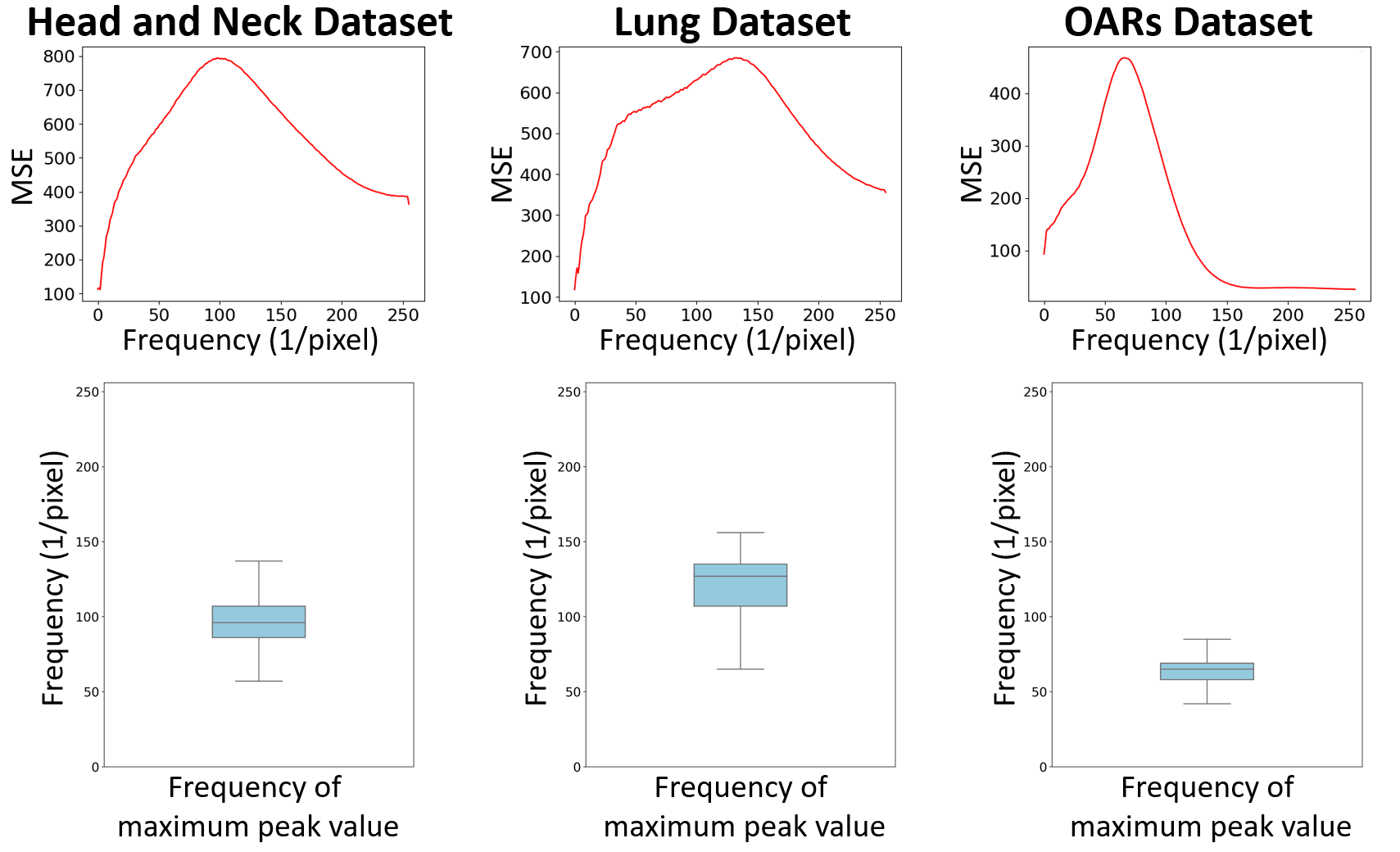}
  \caption{This illustration presents the statistical data of the Mean Squared Error (MSE) of frequency amplitudes between paired CBCT and CT across various spatial frequencies, derived from the frequency domains of three distinct datasets, where the first row is the average MSE distribution for the entire dataset. To further demonstrate that the difference between paired CBCT and CT is mainly concentrated in the intermediate frequencies, we show in the second row the box plot of the distribution of the frequencies corresponding to the maximum peak of MSE for the whole dataset.}
  \label{fig:frequency_analysis}
\end{figure}


\begin{figure*}[ht]
\centering
  \includegraphics[width=.94\textwidth]{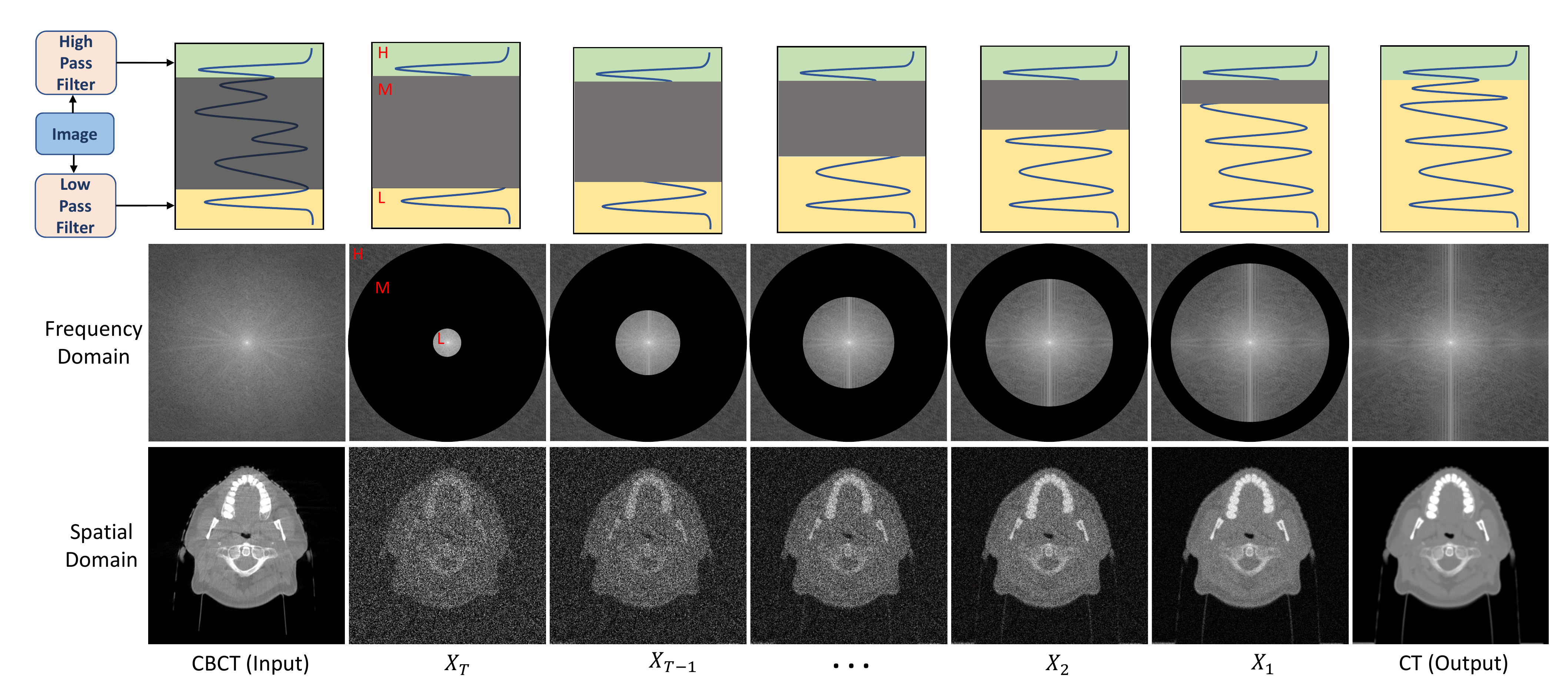}
  \caption{The CBCT-to-CT translation process via the frequency-guided diffusion model (FGDM). The low-frequency information and high-frequency information of CBCTs are used as the diffusion conditions, and the information of intermediate frequencies is gradually generated through the reverse diffusion process to yield translated target images in the CT domain.}
  \label{fig:overview}
\end{figure*}

To the best of our knowledge, we are the first group to implement the diffusion model for zero-shot medical image translation at the anatomical level. The main contributions of this paper are summarised as follows: 

\begin{itemize}
    \item We propose a frequency-guided diffusion model to achieve medical image translation that preserves accurate anatomical information, by using both low-frequency and high-frequency information to guide the generation of intermediate-frequency signals.
    \item We curate and optimize high-pass and low-pass filters to extract the domain invariant information. Specifically, the filtering thresholds of our two filters are freely adjustable at the testing stage to cope with varying domain-invariant zones in the Fourier space between the source and target domains for different types of medical image translation tasks.
    \item Our model achieves the best performance in the CBCT-to-CT translation task, surpassing all other state-of-the-art (SOTA) benchmark models, including GAN-based, VAE-based, and diffusion-based methods. 
\end{itemize}


\section{Related works}
\subsection{Medical Image Translation}

Image-to-image translation has found a wide array of applications in medicine, including tumor and organ segmentation, cross-modal image registration, low-dose CT imaging, fast MR imaging, metal artifact reduction, and treatment planning. For example, Yang et al. \cite{yang2020mri} utilized MRI cross-modal translation to improve segmentation performance by converting T1 MR images into T2 MR images. They found the translated T2 MR images, combined with the original T1 MR images through a multi-channel network, helped to further improve the brain tumor segmentation accuracy. For registration, Fu et al. \cite{fu2020synthetic} translated MR head and neck images into synthetic CT images to serve as a surrogate for MRI-CT registration, which showed reduced target registration errors when compared with direct MRI-CT registration. Medical Image translation has also been applied to low-dose CT imaging denoising \cite{yang2018low}, which translates low-dose CTs to those with image quality equivalent to high-dose CTs. Such an approach helps to reduce the radiation exposure to scanned subjects, while at the same time helping to reduce noise and artifacts present in the original low-dose CT images that may adversely affect the disease diagnosis. MR imaging offers rich information for disease diagnosis and localization. However, MR imaging is usually quite time-consuming and incurs high medical costs. Liu et al. \cite{liu20213d} used medical image translation to generate super-resolution images from coarse images obtained from ultra-fast acquisitions, to save imaging time. Liao et al. \cite{liao2019adn} used a deep learning network to translate metal artifacts-affected CT images to those free of metal artifacts, which significantly improved the CT image quality for further analysis. In image-guided radiation therapy (IGRT), image translation is frequently applied to radiotherapy planning. For instance, CBCT images that are acquired on medical linear accelerators are usually of degraded image quality due to hardware and scanning mode limitations and need to be translated to fan-beam CT-like images to allow accurate radiotherapy planning dose calculation \cite{liang2019generating}. MR imaging is widely used in radiotherapy planning, but the MR planning images also need to be translated to CT images to provide accurate electron density information for dose calculation and treatment planning \cite{kazemifar2019mri}. In general, image-to-image translation is a common need in medicine, which helps to introduce multi-modal information to aid disease diagnosis/localization (segmentation), to harmonize different image modalities for alignment (registration), to improve the quality of available images (low-dose CT denoising, fast MR imaging, metal artifact reduction), and use such improved information to aid treatment planning like radiotherapy planning.

From the perspective of medical image translation principles, the primary idea behind the image-to-image translation task is to establish a mapping between images from two different domains. In medical image translation, the main connection between two domains lies in the underlying anatomical structure information. Maintaining this anatomy during translation is crucial since any deviation or loss can greatly compromise the reliability of the resulting images. To achieve this purpose, the previous works on medical image translation are mainly based on GANs\cite{goodfellow2020generative} and VAEs\cite{kingma2019introduction}. One of the most representative GAN-based methods is CycleGAN\cite{zhu2017unpaired}, which uses cyclic consistency losses to constrain the entire model into two generative networks. It can translate a source domain image to the target domain and vice versa. Another representative work is the geometry-consistent generative adversarial network (GCGAN)\cite{fu2019geometry}, which provides unilateral unsupervised mapping. GCGAN maps the original image into two different predefined geometric transformations and generates two images in a new domain under the corresponding geometric consistency constraints. In addition, RegGAN\cite{kong2021breaking} has been proposed as a new unsupervised method for medical image translation, based on the "loss-correction" theory. It trains a generator with an added registration network to fit misaligned noise distribution, seeking to optimize both image translation and registration tasks. However, GAN-based models have some inherent drawbacks since they are trained by two networks (generator and discriminator) playing against each other, which can render the training difficult and unstable. For VAE-based methods, one of the most representative works on unpaired image translation is UNIT\cite{liu2017unsupervised}, which assumes that two domains share a common latent space and the corresponding images in both domains are mapped to the same latent code. On this basis, MUNIT\cite{huang2018multimodal} is further proposed, which recombines its content code with a random style code taken from the style space of the target domain. However, the general drawback of VAE is that the generated images often suffer from degraded image quality, especially blurriness.

\begin{figure*}[ht]
\centering
  \includegraphics[width=.89\textwidth]{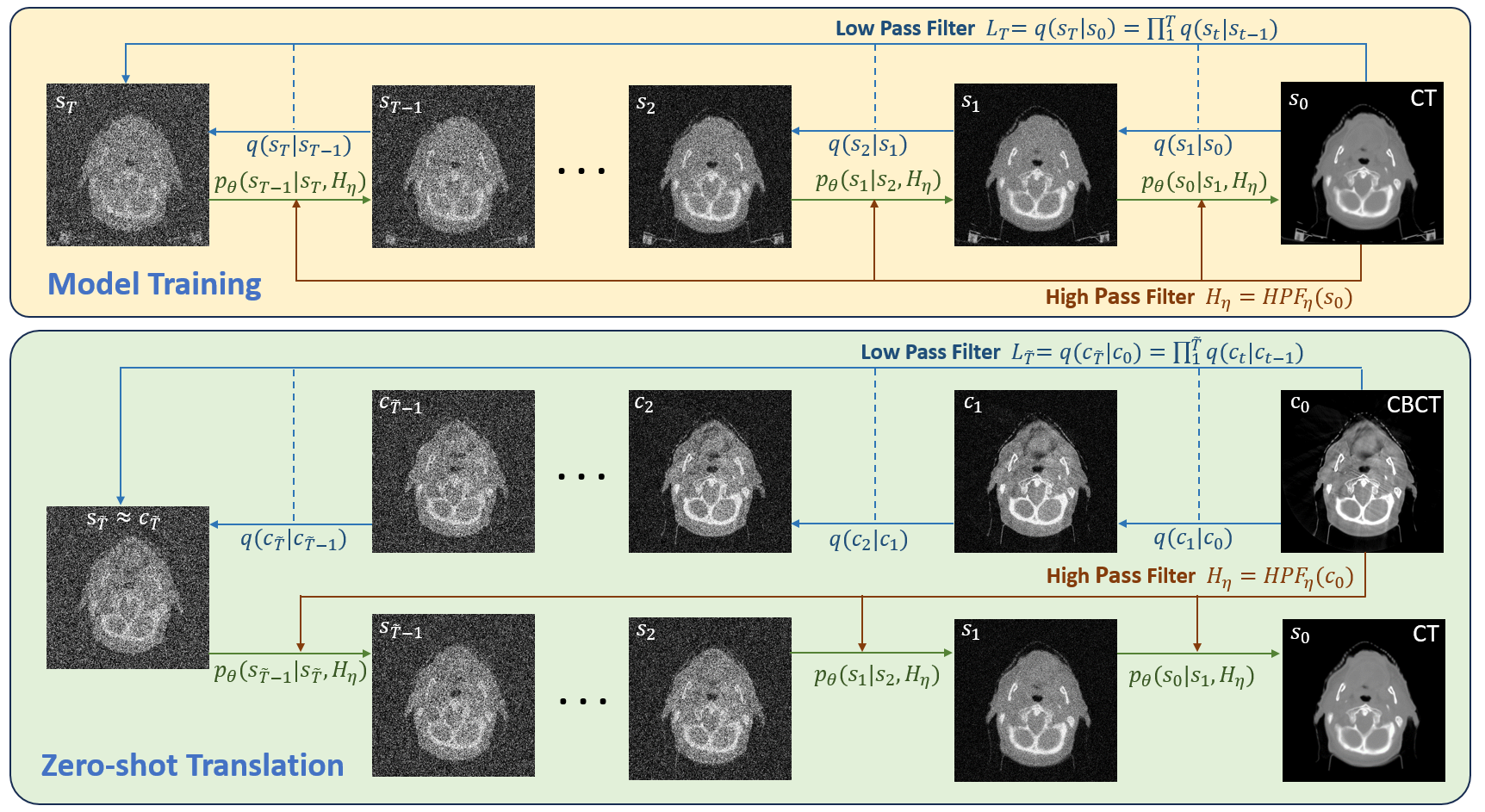}
  \caption{The structure diagram of the frequency-guided diffusion model (FGDM). The model is trained solely on the target image domain (CT) to generate information of the intermediate frequency, conditioned on the low and high frequency CT information. For zero-shot CBCT-to-CT translation, the low and high frequency information of the CBCTs (source) is used instead as conditions to generate corresponding CT images. In the training phase, noise is added until the step $T$, while in the zero-shot translation stage, noise is applied until the step \( \tilde{T} \) (\( \tilde{T} \) $\le$ \( T \)). Using a large $T$ during the training stage is motivated by our desire to have the model adeptly handle a range of low-frequency information thresholds. Using a different \( \tilde{T} \) from $T$ during the zero-shot translation stage allows us to customize the low-frequency information threshold according to the specific disparities between the source and target domains. }
  \label{fig:overview}
\end{figure*}

\subsection{Diffusion Models for Image Translation}
Recently, the latest generative network based on the diffusion model has achieved superior performance in generating high-quality images with excellent realism. Compared with GANs and VAEs, diffusion models also have the ability to achieve zero-shot image translation for their improved robustness to out-of-distribution (OOD) data. One such method is SDEdit\cite{meng2021sdedit}, which translates images by an iterative denoising mechanism via the stochastic differential equation (SDE). SDEdit first adds noise to the source image and subsequently denoises the generated image through the SDE to achieve image translation. Another diffusion model-based translation network is the energy-guided stochastic differential equation (EGSDE)\cite{zhao2022egsde}, which uses pre-trained energy features in the source and target domains to guide unpaired image translation, balancing realism and faithfulness. Based on two feature extractors, EGSDE designs the energy function to promote the retention of domain-independent features and discard domain-specific features. In the application of diffusion modeling for medical image translation, SynDiff\cite{ozbey2023unsupervised} is proposed as a novel adversarial diffusion method for enhanced medical image translation. It uses a conditional diffusion process for accurate image sampling and features a cycle-consistent structure for bilateral modality translation in unpaired datasets. However, SynDiff relies on the translation results of CycleGAN to train the diffusion model, and its performance will be limited by CycleGAN. On the contrary, FGDM does not rely on CycleGAN and is a new zero-shot translation framework. The training of FGDM only requires target domain images, while the training of SynDiff requires inputs from both source and target domains.

When these methods are applied to medical image translation, GAN-based and VAE-based methods tend to achieve better anatomical structure preservation (higher faithfulness) compared to the diffusion model. The diffusion models can achieve higher image quality, i.e., higher realism, but reduced faithfulness\cite{wu2022unifying}. It is thus imperative to improve the diffusion model's faithfulness to the original image, to achieve more accurate medical image translation. Inspired by these challenges, we developed the FGDM framework to retain the high realism of the diffusion model while further improving the corresponding faithfulness.

\section{Methods}
The overview of our model is shown in Fig. \ref{fig:overview}. FGDM can translate unseen source domain data (CBCT) to the target domain data (CT)\cite{liang2019generating}. Low-pass filtering and high-pass filtering of CBCTs are performed to obtain low-frequency information and high-frequency information as diffusion conditions, and the information at intermediate frequencies is gradually generated through a reverse diffusion process to yield final CT images.

\subsection{Diffusion Models }
In diffusion models \cite{sohl2015deep,ho2020denoising,song2021denoising}, there is a forward diffusion process $q(\cdot)$ that gradually adds white Gaussian noise with constant variance $\sigma^2$ to the original data $s_{0}$ in $T$ steps:

\begin{align}\label{eq: forward process1}
    q(s_{1: T} | s_{0})=\prod_{t=1}^{T} q(s_{t} | s_{t-1}) 
\end{align}

\begin{align}\label{eq: forward process2}
 \quad q(s_{t} | s_{t-1})=\mathcal{N}(s_t;\sqrt{\frac{\alpha_t}{\alpha_{t-1}}} s_{t-1}, (1-\frac{\alpha_t}{\alpha_{t-1}}) \sigma^2)
\end{align}
where  $\alpha_{1:t} \in (0, 1]^T$ is a decreasing sequence, and the reverse denoising process $p_{\theta}(\cdot)$ is defined by:

\begin{align}\label{reverse process}
    p_{\theta}(s_{0: T})=p(s_{T}) \prod_{t=1}^{T} p_{\theta}(s_{t-1} | s_{t})
\end{align}

\begin{align}\label{reverse process}
    \quad p_{\theta}(s_{t-1} | s_{t})=\mathcal{N}(s_{t-1} ; \boldsymbol{\mu}_{\theta}(s_{t}, t), {\Sigma}_{\theta}(s_t,t))
\end{align}
where $\boldsymbol{\mu}_{\theta}(s_{t}, t)$ and ${\Sigma}_{\theta}(s_t,t)$ are the mean and variance of the denoising model and $\theta$ denotes its parameters. As shown, the forward diffusion process gradually corrupts the image with introduced noises, which leads to structural detail loss that cannot be recovered by conventional diffusion models.

\subsection{Low Pass Filter}
In this section, we will show that the forward diffusion process can be viewed as a low-pass filter (LPF). First, we set $s_t(x, y), i(x, y),$ and $z(x, y)$ as the degraded image (at step $t$), the original image, and the white Gaussian noise, respectively. Following Ho et al.\cite{ho2020denoising}, we can consider the forward diffusion process as a Markov chain with Gaussian transitions parameterized by the decreasing sequence $\alpha_{1:t} \in (0, 1]^T$. The detailed formulas for $\alpha_{t}$ and $t$ can be found in \cite{xiaotackling, song2020score}, where a special property of the forward process can be derived from Eq. 2 as:  
\begin{equation} \label{e_func}
\begin{aligned}
q(s_t | s_0) = \int q(s_{1:t} | s_0) d s_{1:(t-1)}= \mathcal{N}(s_t; \sqrt{\alpha_t} s_0, (1 - \alpha_t) \sigma^2 )
\end{aligned}
\end{equation}
Correspondingly, we can express $s_t$ as a linear combination of $s_0$ and a noise variable $z$:

\begin{equation} \label{e_func}
\begin{aligned}
s_t = \sqrt{\alpha_t} s_0 + \sqrt{1 - \alpha_t} z, \quad \text{where} \quad z \sim \mathcal{N}(0, \sigma^2)
\end{aligned}
\end{equation}
when $t=0$, $s_0=i$.
\begin{equation} \label{e_func_8}
\begin{aligned}
s_t = \sqrt{\alpha_t} i + \sqrt{1 - \alpha_t} z
\end{aligned}
\end{equation}
In the frequency domain, it can be expressed as:

\begin{equation} \label{e_func}
\begin{aligned}
S_t(u,v)=\mathcal{F}[s_t(x,y)] 
\end{aligned}
\end{equation}
\begin{equation} \label{e_func}
\begin{aligned}
I(u,v)=\mathcal{F}[i(x,y)] 
\end{aligned}
\end{equation}
\begin{equation} \label{e_func}
\begin{aligned}
Z(u,v)=\mathcal{F}[z(x,y)] 
\end{aligned}
\end{equation}
\begin{equation} \label{e_func}
\begin{aligned}
S_t = \sqrt{\alpha_t} I + \sqrt{1 - \alpha_t} Z
\end{aligned}
\end{equation}
Where $S_t(u, v)$, $I(u, v)$, and $Z(u, v)$ are the Fourier transforms of $s_t(x, y)$, $i(x, y)$, and $z(x, y)$, respectively. The autocorrelation function (ACF) of white Gaussian noise can be defined as\cite{nounou2000multiscale}:
\begin{equation} \label{e_func}
\begin{aligned}
 R_N(\tau_1, \tau_2) = \int  Z(u,v)Z(u-\tau_1,v-\tau_2)du dv 
\end{aligned}
\end{equation}
where $\tau_1$ and $\tau_2$ denote lags.
In the case of white Gaussian noise, the signal $Z(u,v)$ is a random variable with zero mean and constant variance $\sigma^2$:
\begin{equation} \label{e_func}
\begin{aligned}
E[Z(u,v)] = 0 
\end{aligned}
\end{equation}

\begin{equation} \label{e_func}
\begin{aligned}
 Var[Z(u,v)] = \sigma^2 
\end{aligned}
\end{equation}
Therefore, the autocorrelation function of white Gaussian noise is:

\begin{equation} \label{e_func}
\begin{aligned}
 R_N(\tau_1,\tau_2) =  \int \sigma^2 \delta(\tau_1,\tau_2)dudv = \sigma^2 \delta(\tau_1,\tau_2)
\end{aligned}
\end{equation}
Where $\delta(\tau_1,\tau_2)$ is the Dirac delta function. Therefore, the power of the white Gaussian noise is the same across all frequencies.

\begin{equation} \label{e_func}
\begin{aligned}
PSD(u,v)_Z={\sigma}^{2} \ \ for\  any\  (u,v)
\end{aligned}
\end{equation}
In general images, the power spectral density (PSD) is related to the spatial frequency as follows\cite{van1996modelling,tolhurst1992amplitude}:
\begin{equation} \label{e_func}
\begin{aligned}
PSD(u,v)_I=\frac{k}{\Vert(u,v)\Vert^a} \ \ (k>0,a>1)
\end{aligned}
\end{equation}
where $k$ and $a$ are scaling/modifying factors, which indicates that $PSD(u,v)_I \propto \frac{1}{\Vert(u,v)\Vert}$. When white Gaussian noise is added to an image, the PSD of the noise is added to the PSD of the image. If $PSD_{I}$ is the power of the original image and $PSD_{Z}$ is the power of the added white Gaussian noise, the signal-to-noise ratio (SNR) in step $t$ can be defined as:

\begin{equation} \label{e_func}
\begin{aligned}
SNR_{t} = \frac{ \sqrt{\alpha_t} PSD_I  }{ \sqrt{1-\alpha_t}  PSD_{Z}}
\end{aligned}
\end{equation}

\begin{equation} \label{e_func}
\begin{aligned}
SNR_{t} = \frac{\sqrt{\alpha_t} k }{\sqrt{1-\alpha_t} \Vert(u,v)\Vert^a \sigma^2} \ \ (k>0,a>1)
\end{aligned}
\end{equation}
From Eq. 19, the $SNR_{t}$ is increasing with $\alpha_t$. Since $\alpha_t$ is a decreasing sequence ($\alpha_t \propto \frac{1}{t}$), the relation between $SNR_{t}$ and $t$ can be established as:

\begin{equation} \label{e_func}
\begin{aligned}
SNR_{t} \propto  \frac{1}{t}
\end{aligned}
\end{equation}
From Eq. 19, the $SNR_{t}$ at any frequency $\Vert(u,v)\Vert$ is also directly proportional to the power of the original signal at that frequency, and is decreasing with increasing frequency. Thus, with more forward diffusion steps  $t$, $SNR_{t}$ decreases and the signals at higher frequencies are earlier-to-be impacted. Considering $\phi$ as the SNR threshold below which the information is completely corrupted, and depending on the desired frequency threshold $\psi$ of our low-pass filter, there exists an appropriate number of steps $t$ beyond which all the non-low-frequency information will be lost:

\begin{equation} \label{e_func}
SNR_{t} \left\{
\begin{aligned}
< & \phi  \qquad  \Vert(u,v)\Vert>\psi   \\
\geq & \phi  \qquad \Vert(u,v)\Vert \leq \psi  \\
\end{aligned}
\right.
\end{equation}
If we denoted this step $t$ as $\Tilde{T}$, the forward diffusion equals to low-pass filter:

\begin{equation} \label{e_func}
\begin{aligned}
L_{\Tilde{T}} = LPF_{\Tilde{T}}(i(x,y))=  q(s_{\Tilde{T}} | s_0) 
\end{aligned}
\end{equation}

\subsection{High Pass Filter}
We use the Sobel operator\cite{kanopoulos1988design} as our high-pass filter (HPF), which is technically a discrete differential operator that computes the gradient of the image intensity function to obtain the high-frequency information. Specifically, we first convolve the image with the horizontal and vertical Sobel filter kernels ${K}_x$ and ${K}_y$, respectively, to obtain the filtered images ${G}_x$ and ${G}_y$. At each point in the image, the resulting gradient approximations can be combined to give the gradient magnitude using the following equations:
\begin{equation} \label{e_func}
\begin{aligned}
G_{x} = K_x \circledast i,\  G_{y} = K_y \circledast i
\end{aligned}
\end{equation}

\begin{equation} \label{e_func}
\begin{aligned}
i_{mag} = \sqrt{G_x^2 + G_y^2} 
\end{aligned}
\end{equation}
The high-frequency information $H$ is then obtained through the threshold $\eta$:
\begin{equation} \label{e_func}
H_{\eta} = HPF_{\eta}(i(x,y)) = \left\{
\begin{aligned}
 &i_{mag}(x,y)  \qquad   &i_{mag}(x,y)\geq \eta   \\
 &0  \qquad &i_{mag}(x,y)<\eta   \\
\end{aligned}
\right.
\end{equation}

\subsection{Model Training}
In our model, we choose a hybrid diffusion model, denoised diffusion GAN\cite{xiaotackling}, as the backbone network, which employs a special f-divergence instance called softened reverse Kullback–Leibler divergence (SRKL) \cite{shannon2020non}. It is able to sample the image in less than 8 steps to substantially accelerate the inference speed, while the original DDPM requires over 1000 steps. It can achieve a performance close to the original DDPM at an inference speed that is hundreds of times faster than the original DDPM\cite{xiaotackling}. The training is formulated by matching the conditional denoising model $p_{\theta}(s_{t-1}|s_t, H_{\eta})$ and $q(s_{t-1}|s_t)$ using an adversarial loss that minimizes a divergence $D_{adv}$ per denoising step:

\begin{equation} \label{e_func}
\begin{aligned}
\mathop{\min}_{\theta} \sum_{t \geq 1} \mathbb{E}_{q(x_t)}[D_{adv}(q(s_{t-1}|s_t)||p_{\theta}(s_{t-1}| s_t, H_{\eta}))]
\end{aligned}
\end{equation}
The SRKL is a time-dependent discriminator, and it is trained through the following equation: 

\begin{equation} \label{e_func}
\begin{aligned}
\mathop{\min}_{\rho} \sum_{t \geq 1} \mathbb{E}_{q(s_t)}[\mathbb{E}_{q(s_{t-1}|s_t)}[-log(SRKL_{\rho}(s_{t-1},s_t,t)+\\ \mathbb{E}_{p_{\theta}(s_{t-1}|s_t,H_{\eta})}[-log(1-SRKL_{\rho}(s_{t-1},s_t,H_{\eta},t) )]]
\end{aligned}
\end{equation}
Given the SRKL, we train the conditioned denoising diffusion model by:
\begin{equation} \label{e_func}
\begin{aligned}
\mathop{\max}_{\theta} \sum_{t \geq 1} \mathbb{E}_{q(s_t)}\mathbb{E}{p_{\theta}(s_{t-1}|s_t,H_{\eta})}[log(SRKL_{\rho}(s_{t-1}; s_t; H_{\eta}; t))]
\end{aligned}
\end{equation}

\subsection{Zero-Shot Image Translation}
In zero-shot image translation, the model is trained only on the target domain data at the training time, as the images from the source domain are not available during training. The model needs to respond directly to samples from domains that were not observed during training. Zero-shot approaches usually associate observed and unobserved classes through some form of auxiliary information, while in the CBCT-to-CT image translation task, we found that their low- and high-frequency information is approximately the same, with the main difference focusing on the information at the intermediate frequencies. The image translation task between two modalities/domains can then be described as Eq. \ref{e_29}, Eq. \ref{e_30}, and Algorithm \ref{alg;solver}:

\begin{equation} \label{e_29}
\begin{aligned}
q(s_{\Tilde{T}} | c_0) = q(c_{\Tilde{T}} | c_0)= \mathcal{N}(s_{\Tilde{T}}; \sqrt{\alpha_{\Tilde{T}}} c_0, (1 - \alpha_{\Tilde{T}}) \sigma^2 )
\end{aligned}
\end{equation}

\begin{equation} \label{e_30}
\begin{aligned}
p_{\theta}(s_0|s_t ) =& \prod_{t = 1}^{\Tilde{T}} p_{\theta}(s_{t-1}| s_t ,HPF_{\eta}(c_0) )
\end{aligned}
\end{equation}

\begin{algorithm}
\caption{Frequency Guided Zero-Shot Translation}
\label{alg;solver}
\SetAlgoLined
\SetKwInput{KwData}{Inputs}
\SetKwInput{KwResult}{Output}
\KwData{The source image $c_0$, Pre-trained denoising diffusion model $p_{\theta}$, Step of forward diffusion ${\Tilde{T}}$, Low pass filter $LPF(\cdot)$, High pass filter $HPF(\cdot)$, Constant variance $\sigma^2$, Decreasing sequence $\alpha_{1:t} \in (0, 1]^T$;}
\KwResult{Generated high-faithfulness translated image $s_0$;}
 Initialize all parameters and variables\;
\begin{algorithmic}
    \STATE $z \sim \mathcal{N}(0, \sigma^2)$
    \STATE $L_{\Tilde{T}} \gets LPF_{\Tilde{T}}(c_0) = \sqrt{\alpha_{\Tilde{T}}} c_0 + \sqrt{1 - \alpha_{\Tilde{T}}} z $ 
    \STATE $H_{\eta} \gets HPF_{\eta}(c_0)$
    \STATE $s_{t} \gets L_{\Tilde{T}}$
    \FOR{$t = \Tilde{T} $ to $1$}
        \STATE $s_{t-1} \gets p_{\theta}(s_{t-1}| s_t ,H_{\eta}) $
        \STATE $s_{t} \gets s_{t-1}$
    \ENDFOR
    \STATE $s_{0} \gets s_{t}$
    \RETURN $s_0$
\end{algorithmic}
\end{algorithm}


\section{Experiments and Discussions}

\subsection{Datasets and pre-processing}
\subsubsection{Head and Neck Dataset}
We trained and tested the FGDM on the head and neck dataset. The head and neck dataset contained patient-specific CT and CBCT images divided into 70 training, 9 validation, and 20 testing cases, and the validation data was used to select the best performance checkpoint. The CBCT images had a voxel spacing of 0.51 × 0.51 × 1.99 $mm^3$ and the voxel spacing of CT images was 1.17 × 1.17 × 3.00 $mm^3$. All the images were intercepted by HU values [-1000,1000], and were linearly normalized to [0,1]. We used the images with 2D slices as input, with all slices resized to 192 × 192. For testing, an open-source deformable image registration toolbox\cite{klein2009elastix} was used to align the CT to CBCT, and the registered CT images were used as the reference for assessing the performance of image translation.

\subsubsection{Lung Dataset}
To further test the zero-shot capability of our model, we applied the model trained on the head and neck CTs directly to the lung CBCT-to-CT translation task. In total, we collected lung CBCT and CT images from 18 patients, which underwent the same pre-processing steps as the head and neck data, including HU value interception, linear normalization, and slice resizing. Similar to the head and neck dataset, the lung CTs were also deformably registered to the corresponding CBCTs for each patient to serve as the evaluation references.

\subsubsection{Organs at Risk (OARs) Dataset}
To rigorously evaluate the performance under challenging and varied conditions, we utilized the Organs at Risk (OARs) dataset from the American Association of Physicists in  Medicine (AAPM) thoracic auto-segmentation grand challenge\cite{dahiya2021multitask,yang2018autosegmentation}. This dataset came from 24 patients and it incorporated paired CBCTs and CTs of anatomies including the esophagus, spinal cord, heart, and lungs. The OARs dataset was pre-processed similarly as the head and neck dataset and the lung dataset, which involved intercepting the HU values to the range of [-1000, 1000], followed by a linear normalization to [0,1]. The 2D slices were used as input, with all slices resized to 192 × 192. Including this dataset in evaluation helps to further assess the robustness and versatility of FGDM in scenarios with more variations in data distribution.

\subsubsection{Head MR T1 Dataset}

Another dataset utilized in our assessment consists of Head MR T1 images acquired from two different repositories (centers): The Neurofeedback Skull-stripped (NFBS)\footnote{\url{http://preprocessed-connectomes-project.org/NFB_skullstripped}} repository and the MICCAI 2020 challenge: Anatomical Brain Barrier to Cancer Spread (ABCs)\footnote{\url{https://abcs.mgh.harvard.edu/index.php}} repository. This dataset was chosen to reflect real-world needs in cross-center image translation, where data from different institutions can vary due to differing imaging protocols and need to be harmonized for collective analysis. The NFBS dataset contained structural T1-weighted and anonymized (de-faced) images from 125 participants. This collection served as the source domain data for FGDM training. The ABCs dataset comprised 60 patients diagnosed with glioblastoma and low-grade glioma, which served as the target domain data for testing. For pre-processing, we clipped the image intensity between the $0^{th}$ and $99^{th}$ percentiles to remove outliers and standardize the dynamic range of the image. We then normalized the images to [0,1] and resized all slices to 256 × 256 in dimension.

\begin{figure*}[]
\centering
  \includegraphics[width=.99\textwidth]{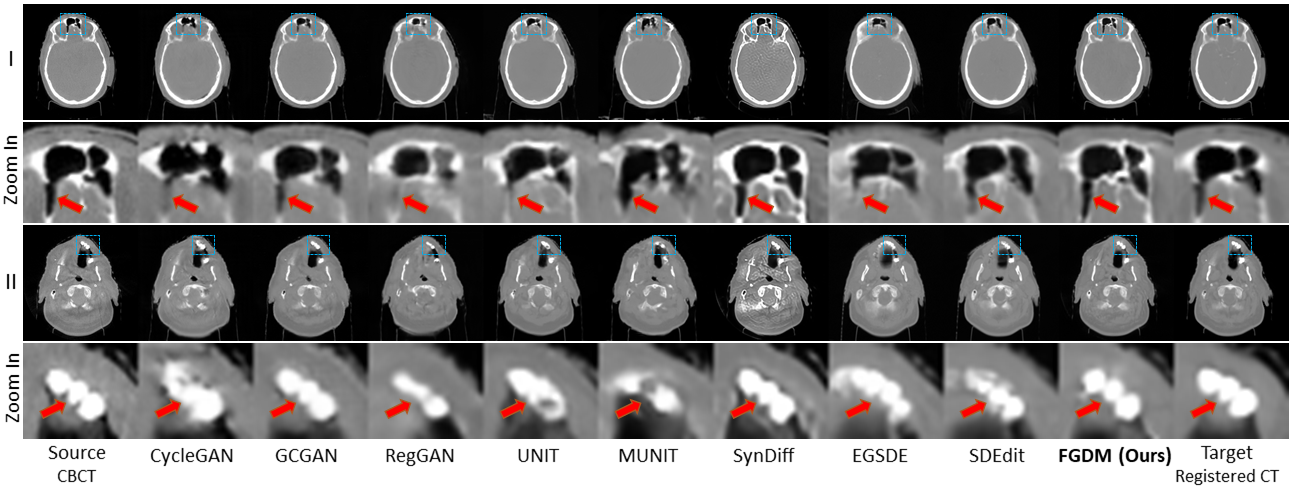}
  \caption{Visual comparison between GAN-based, VAE-based, and diffusion model-based CBCT-to-CT translation methods on the head and neck dataset (in-distribution). The display window is [-1000,1000] HU.}
  \label{fig:head and neck}
\end{figure*}

\begin{table*}[ht]
\centering
\caption{Quantitative comparison between GAN-based, VAE-based, and diffusion model-based CBCT-to-CT translation methods on the head and neck dataset (in-distribution). In the table, the mean values calculated on the test set were presented for all metrics, and the arrows are pointing to the directions of improved accuracy.}
\renewcommand\arraystretch{1.1}
\setlength{\tabcolsep}{4.5mm}
\begin{tabular}{c|c|cc|ccc}
\Xhline{1pt}
\multirow{2}{*}{Model}             & \multirow{2}{*}{FID $\downarrow$   } & \multicolumn{2}{c|}{With Source} & \multicolumn{3}{c}{With Target} \\ \cline{3-7} 
         &             & PSNR $\uparrow$ & SSIM $\uparrow$    & PSNR $\uparrow$ & SSIM $\uparrow$ & Frequency MSE $\downarrow$ \\ \hline
CycleGAN & 47.6 & 22.5 & 0.844  & 22.6 & 0.867& 440\\
GCGAN & 42.3  & 23.6 & 0.863 & 23.7 & 0.886& 415\\ 
RegGAN &  40.5 & 19.9 & 0.765 & 20.6 & 0.793&  442 \\ 
UNIT & 36.6  & 22.2 & 0.821  & 22.8 & 0.854&  436\\
MUNIT & 44.5  & 21.2 & 0.811  & 21.9 & 0.856&457\\  
SynDiff  &  49.0 & 26.0 & 0.902 & 22.9 & 0.862 & 502 \\
EGSDE &41.9  & 22.6 & 0.800  & 23.4 & 0.837& 461\\
SDEdit & 33.2  & 24.8 & 0.830 & 25.0 & 0.857& 422\\
\textbf{FGDM (Ours)} &   \textbf{24.0}  & \textbf{27.8} & \textbf{0.908} & \textbf{26.4} & \textbf{0.904} & \textbf{400} \\ \Xhline{1pt}
\end{tabular}
\label{tab:head and neck}
\end{table*}

\begin{figure*}[ht]
\centering
  \includegraphics[width=.99\textwidth]{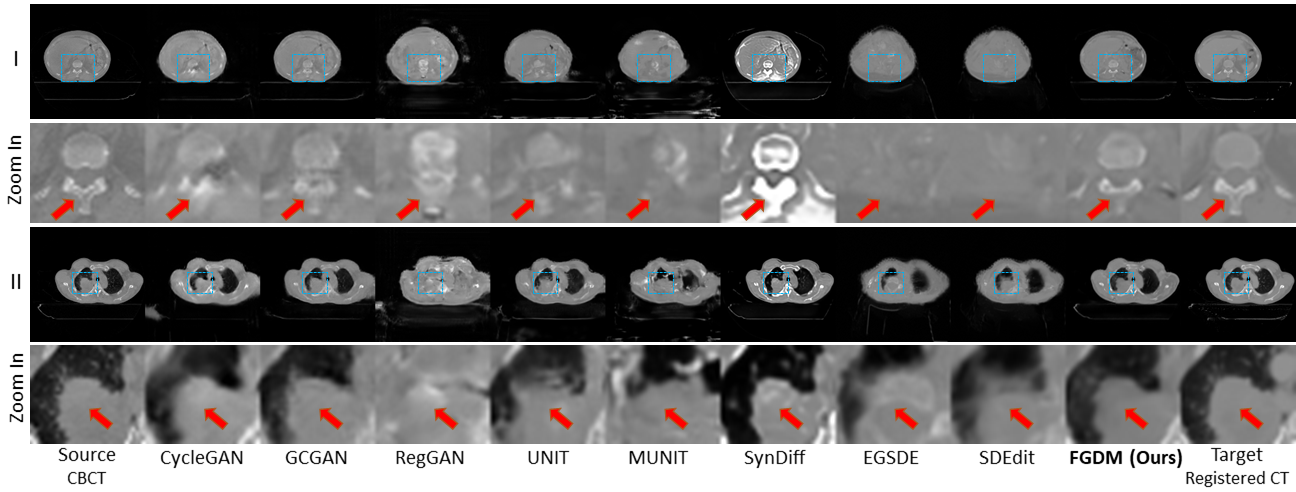}
    \caption{Visual comparison between GAN-based, VAE-based, and diffusion model-based CBCT-to-CT translation methods on the lung dataset (out-of-distribution). The display window is [-1000,1000] HU.}
  \label{fig:lung}
\end{figure*}

\begin{table*}[ht]
\centering
\caption{Quantitative comparison between GAN-based, VAE-based, and diffusion model-based CBCT-to-CT translation methods on the lung dataset (out-of-distribution). In the table, the mean values calculated on the test set were presented for all metrics, and the arrows are pointing to the directions of improved accuracy.}
\renewcommand\arraystretch{1.1}
\setlength{\tabcolsep}{4.5mm}
\begin{tabular}{c|c|cc|ccc}
\Xhline{1pt}
\multirow{2}{*}{Model}             & \multirow{2}{*}{FID $\downarrow$   } & \multicolumn{2}{c|}{With Source} & \multicolumn{3}{c}{With Target} \\ \cline{3-7} 
         &             & PSNR $\uparrow$ & SSIM $\uparrow$    & PSNR $\uparrow$ & SSIM $\uparrow$ & Frequency MSE $\downarrow$ \\ \hline 
CycleGAN & 62.5  & 21.8 & 0.838 & 22.1 & 0.838 & 454 \\
GCGAN & 59.6  & 23.5 & 0.876 & 23.4 & 0.868 & 426 \\ 
RegGAN  &  86.1 & 16.9 & 0.66 & 17.2 & 0.67 & 503 \\
UNIT & 69.7  & 21.6 & 0.776  & 21.7 & 0.785  & 461\\
MUNIT & 90.8  & 20.3 & 0.759  & 20.5 & 0.771  & 501\\  
SynDiff  &  60.8 & 24.2 & 0.872 & 23.1 & 0.839 & 514 \\
EGSDE &101.0  & 23.6 & 0.799  & 23.2 & 0.812  & 505\\
SDEdit &81.5  & 25.7 & 0.826  & 25.0 & 0.836  & 460\\
\textbf{FGDM (Ours)} & \textbf{42.4}  & \textbf{28.4} & \textbf{0.906} & \textbf{27.4} & \textbf{0.901}  & \textbf{413}\\ \Xhline{1pt}
\end{tabular}
\label{tab:lung}
\end{table*}

\begin{figure*}[ht]
\centering
  \includegraphics[width=.99\textwidth]{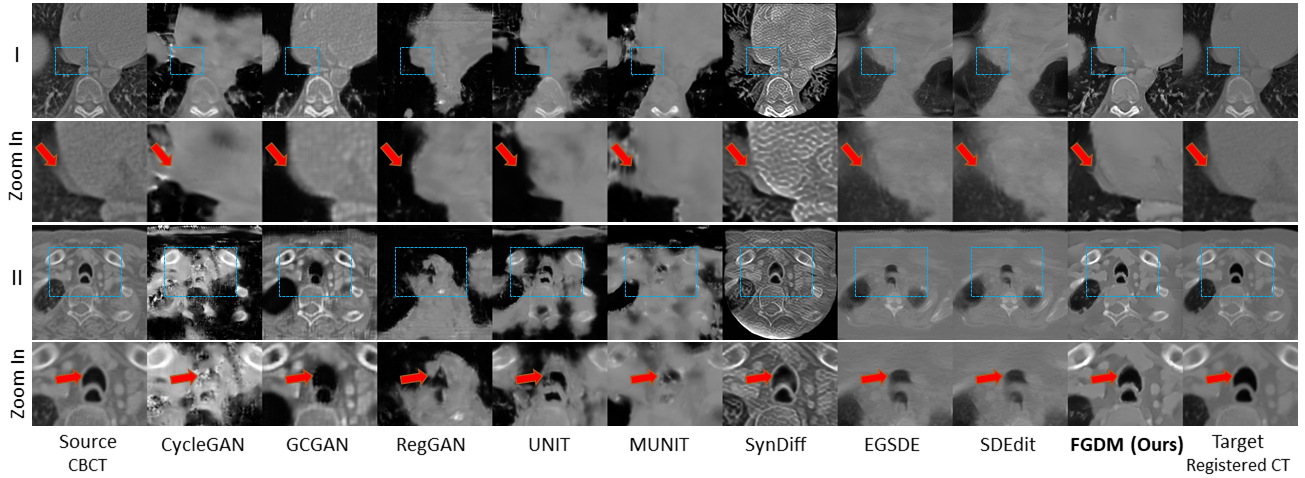}
    \caption{Visual comparison between GAN-based, VAE-based, and diffusion model-based CBCT-to-CT translation methods on the OARs dataset (out-of-distribution). The display window is [-1000,1000] HU.}
  \label{fig:aapm_result}
\end{figure*}

\begin{table*}[ht]
\centering
\caption{Quantitative comparison between GAN-based, VAE-based, and diffusion model-based CBCT-to-CT translation methods on the OARs dataset (out-of-distribution). In the table, the mean values calculated on the test set were presented for all metrics, and the arrows are pointing to the directions of improved accuracy.}
\renewcommand\arraystretch{1.1}
\setlength{\tabcolsep}{4.5mm}
\begin{tabular}{c|c|cc|ccc}
\Xhline{1pt}
\multirow{2}{*}{Model}             & \multirow{2}{*}{FID $\downarrow$   } & \multicolumn{2}{c|}{With Source} & \multicolumn{3}{c}{With Target} \\ \cline{3-7} 
         &             & PSNR $\uparrow$ & SSIM $\uparrow$    & PSNR $\uparrow$ & SSIM $\uparrow$ & Frequency MSE $\downarrow$ \\ \hline 
CycleGAN  &  118.0 & 14.5 & 0.554 & 14.8 & 0.573 & 294\\
GCGAN  &  142.0 & 17.5 & 0.724 & 17.6 & 0.751 & 216\\
RegGAN  &  271.0 & 12.4 & 0.412 & 13.1 & 0.459 & 316\\
UNIT  &  158.0 & 14.4 & 0.562 & 15.0 & 0.612 & 274\\
MUNIT  &  204.0 & 16.6 & 0.616 & 17.3 & 0.653 & 282\\
SynDiff  &  184.0 & 15.6 & 0.589 & 15.5 & 0.558 & 520\\
EGSDE  &  198.0 & 21.1 & 0.816 & 21.4 & 0.835 & 306\\
SDEdit  &  191.0 & 20.2 & 0.79 & 21.0 & 0.832 & 301\\
\textbf{FGDM (Ours)}  &  \textbf{94.8} & \textbf{24.0} & \textbf{0.857} & \textbf{23.0} & \textbf{0.889}  & \textbf{176}\\ \Xhline{1pt}
\end{tabular}
\label{tab:aapm_result}
\end{table*}

\begin{figure*}[ht]
\centering
  \includegraphics[width=.99\textwidth]{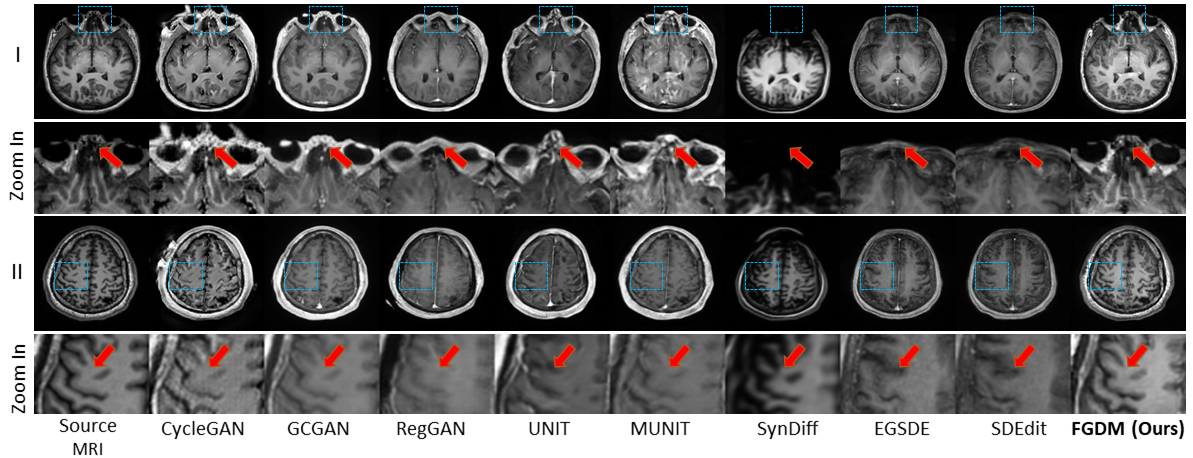}
    \caption{Visual comparison between GAN-based, VAE-based, and diffusion model-based head MR T1 translation methods on the head MRI dataset.}
  \label{fig:mrt1_result}
\end{figure*}

\begin{table*}[ht]
\centering
\caption{Quantitative comparison between GAN-based, VAE-based, and diffusion model-based methods on the head T1 MR dataset. In the table, the mean values calculated on the test set were presented for all metrics, and the arrows are pointing to the directions of improved accuracy.}
\renewcommand\arraystretch{1.1}
\setlength{\tabcolsep}{4.5mm}
\begin{tabular}{c|c|cc}
\Xhline{1pt}
\multirow{2}{*}{Model}             & \multirow{2}{*}{FID $\downarrow$   } & \multicolumn{2}{c}{With Source} \\ \cline{3-4} 
         &   & PSNR $\uparrow$ & SSIM $\uparrow$ \\ \hline
CycleGAN  &  86.4 & 16.1 & 0.664  \\
GCGAN  &  61.2 & 16.6 & 0.706  \\
RegGAN  &  76.4 & 14.7 & 0.616  \\
UNIT  &  66.0 & 14.0 & 0.569  \\
MUNIT  &  66.6 & 17.0 & 0.708  \\
SynDiff  &  104.0 & 16.7 & 0.693 \\
EGSDE  &  69.1 & 19.8 & 0.687  \\
SDEdit  &  61.5 & 20.0 & 0.656  \\
\textbf{FGDM (Ours)} &  \textbf{59.3} & \textbf{21.2} & \textbf{0.805} \\
 \Xhline{1pt}
\end{tabular}
\label{tab:mrt1}
\end{table*}

\begin{table*}[ht]
\centering
\caption{Ablation study of low and high frequency information. The arrows are pointing to the directions of improved accuracy.}
\renewcommand\arraystretch{1.1}
\setlength{\tabcolsep}{4.5mm}
\begin{tabular}{c|c|cc|cc}
\Xhline{1pt}
\multirow{2}{*}{High Frequency Ratio}   & \multirow{2}{*}{FID$\downarrow$   } & \multicolumn{2}{c|}{With Source} & \multicolumn{2}{c}{With Target} \\ \cline{3-6} 
         &      & PSNR $\uparrow$ & SSIM $\uparrow$  & PSNR $\uparrow$ & SSIM $\uparrow$ \\ \hline

FGDM w/o low-freq info guidance &30.2 & 24.3 & 0.89 & 23.8 & 0.889 \\ 
FGDM w/o high-freq info guidance & 38.2 & 21.5 & 0.748  & 21.8 & 0.778\\
\textbf{FGDM (Ours)} &   \textbf{24.0} & \textbf{27.8} & \textbf{0.908} & \textbf{26.4} & \textbf{0.904} \\ \Xhline{1pt}
\end{tabular}
\label{tab:abalation}
\end{table*}

\subsection{Implementation Details}
We trained our model via the standard Adam optimizer, using an initial learning rate of $10^{-4}$ under the Cosine Annealing learning rate scheduler, by setting the minimum learning rate to $10^{-5}$. The batch size was set to 8, for a training of 200 epochs. In our model, the signal extractions at low and high frequencies are controlled by the forward diffusion step number $\Tilde{T}$ and the Sobel filtering threshold $\eta$ (Eq. 25), respectively. To train a model that is robust to the variations of domain-invariant zones between the source and target images in the frequency space, we randomized $\eta$ during the training by values from 1-25 to extract different levels of high-frequency signals for conditioning. For low-frequency signals, such a variation has already been built in the stepwise forward diffusion process, and we can directly perform test-time evaluation by changing the forward diffusion step number on the testing source-domain image.  In this study, we choose $\eta$ as 10 and  $\Tilde{T}$ as 4 for our testing set, after performing a parameterized study of the two parameters in the ablation study section. We would like to point out that such a parameterized study is directly conducted at the testing stage due to the unique design of FGDM, which is highly adaptive and adjustable to different data inputs without requiring any model re-training.
All networks were implemented using the Pytorch library and we ran the experiments on an NVIDIA Tesla V100 GPU. All methods implemented for comparison used their official open source codes\cite{zhu2017unpaired,fu2019geometry,liu2017unsupervised,huang2018multimodal,zhao2022egsde,meng2021sdedit}, where SDEdit and EGSDE started with a step count of 200.
Our code is available at \href{https://github.com/Kent0n-Li/FGDM}{https://github.com/Kent0n-Li/FGDM}.

\subsection{Comparison Metrics and Methods}
To compare the image translation performance of different models, we measured both realism and faithfulness of the translated images. For realism, we report the widely used Frechet Inception Distance Score (FID) \cite{heusel2017gans} between the translated images and the target reference images. Considering that our target reference images are obtained by deformable registration and are not exactly ground truth images due to non-deformation-induced changes and deformable registration errors, we evaluated faithfulness by comparing our translated images to both the source CBCT images and the reference deformed CT images, as it is critical to evaluate the faithfulness of the structure retention from the source CBCT images. Such a strategy has been used by many works\cite{park2020contrastive,zhao2022egsde,meng2021sdedit} that do not have ground truth images to calculate faithfulness. For faithfulness, we report PSNR (peak signal-to-noise ratio), and structural similarity index measure (SSIM)\cite{wang2004image} of the generated images with both the source and target images. In addition, to further validate the performance of FGDM, we also calculated the MSE (mean squared error) in the spatial frequency domain between the generated images and the target images.
For a comprehensive comparison with other image translation methods, we evaluated a total of three classes of widely used image translation methods ( GAN-based, VAE-based, and diffusion model-based). Among them, GAN-based and VAE-based methods require both source-domain and target-domain images for training, while the diffusion model-based methods can achieve zero-shot, e.g., SDEdit.

\subsection{Performance on the Head and Neck Dataset}

The widely used frameworks in GAN-based image translation methods include CycleGAN and GCGAN, both of which require images from the source and target domains to be fed into the network during training. From the results shown in  Table \ref{tab:head and neck}, CycleGAN and GCGAN achieve comparatively good scores in SSIM, but perform poorly on FID. In general, both CycleGAN and GCGAN enforce structure retention by adding extra loss functions, which however adversely affect the quality of the images, resulting in a lower FID. As for RegGAN, although its performance on FID is slightly better than CycleGAN, the corresponding performance on SSIM and PSNR is worse. UNIT and MUNIT are VAE-based image translation methods, and they align the latent space of the source and target domains so that they share the same latent space. UNIT outperforms GAN on FID, but performs poorly on SSIM, most likely because UNIT and MUNIT lose structure details during the latent space mapping. SDEdit and EGSDE are both diffusion model-based image translation methods, and SDEdit achieves a performance second only to FGDM on FID. It demonstrates the potential of diffusion models for generating high-quality images with excellent realism. However, both SDEdit and EGSDE lose substantial anatomical details in the forward diffusion, resulting in poor faithfulness as shown by metrics including PSNR and SSIM. SynDiff, another diffusion model-based method, performed poorly with an FID of 49.0, although it had the second-highest SSIM with the source image at 0.902. As for the inference time, the diffusion GAN used by FGDM takes about 0.2 seconds to translate a single image, while SDEdit takes about 9 seconds. Overall, FGDM outperforms other methods in all metrics, as it not only inherits the diffusion model's ability to produce high-quality and realistic images but also preserves anatomical details through frequency-domain guidance. Besides, FGDM performs the best in reducing the MSE for the frequency domain. Since the frequency differences between different domains are not entirely in the medium range but also cover the high- and low-frequency range (Fig. 2), FGDM corrects not only mismatches in the medium frequency range but to some extent also those in the low and high-frequency ranges, demonstrated in some cases of removing high-frequency streak artifacts.

Through the visual comparison in Fig. \ref{fig:head and neck}, the images of SDEdit and EGSDE appear generally acceptable on a coarse scale. However, when we zoom into the structure details, for instance the nasal cavity and teeth regions, the structural loss and distortions can be clearly seen in images of both methods.  As for the results of VAE-based methods in Fig. \ref{fig:head and neck}, the anatomical details are similarly not well preserved. For VAE-based methods, an encoder is first used to convert the image into its latent space, followed by a decoder to generate the translated image from the latent space. Structural information is often lost during the encoding, which leads to blurred images of VAE-based methods. Relatively, GCGAN generates images that better preserve the anatomical details for a more balanced realism and faithfulness, while it is still inferior compared to FGDM. The zoomed image shows that FGDM effectively preserves the structural details and generates images with these details closest to those of the source image, compared to other methods. Although FGDM is already the best performer among all the methods, some small structure mismatches from the source image can still be observed, which warrants future development and optimization.

\subsection{Performance on the Lung Dataset}
A high-performance zero-shot model should demonstrate excellent robustness and generalization on out-of-distribution data. Therefore, we tested the translation performance from CBCT to CT by applying the models trained on the head and neck dataset directly to the lung dataset. As shown in Table \ref{tab:lung} and Fig. \ref{fig:lung}, most of the benchmark models performed poorly on this task, indicating the impact of a shift in testing data distributions. Most methods suffer from the loss of structural details and hallucination structures after the distribution shift without fine-tuning, while FGDM still maintains the accuracy of the anatomical structures for robust zero-shot image translation.

\begin{table*}[ht]
\centering
\caption{Impact of the high-frequency ratio parameter $\eta$. The arrows are pointing to the directions of improved accuracy.}
\renewcommand\arraystretch{1.1}
\setlength{\tabcolsep}{4.5mm}
\begin{tabular}{c|c|cc|cc}
\Xhline{1pt}
\multirow{2}{*}{High Frequency Ratio ($\eta$)}   & \multirow{2}{*}{FID $\downarrow$   } & \multicolumn{2}{c|}{With Source} & \multicolumn{2}{c}{With Target} \\ \cline{3-6} 
         &           & PSNR $\uparrow$ & SSIM $\uparrow$   & PSNR $\uparrow$ & SSIM $\uparrow$ \\ \hline

5 &26.9 & 26.7 & 0.892  & 25.5 & 0.881\\ 

\textbf{10} &  \textbf{24.0}  & \textbf{27.8} & \textbf{0.908}& \textbf{26.4} & \textbf{0.904} \\

15 & 25.1  & 27.7 & 0.902  & 26.4 & 0.903  \\  
20 & 30.1  & 26.8 & 0.881  & 26.2 & 0.89 \\ 
25 &34.6 & 26.4 & 0.869  & 25.9 & 0.88\\ \Xhline{1pt}
\end{tabular}
\label{tab:eta}
\end{table*}							

\begin{figure*}[ht]
\centering
  \includegraphics[width=.85\textwidth]{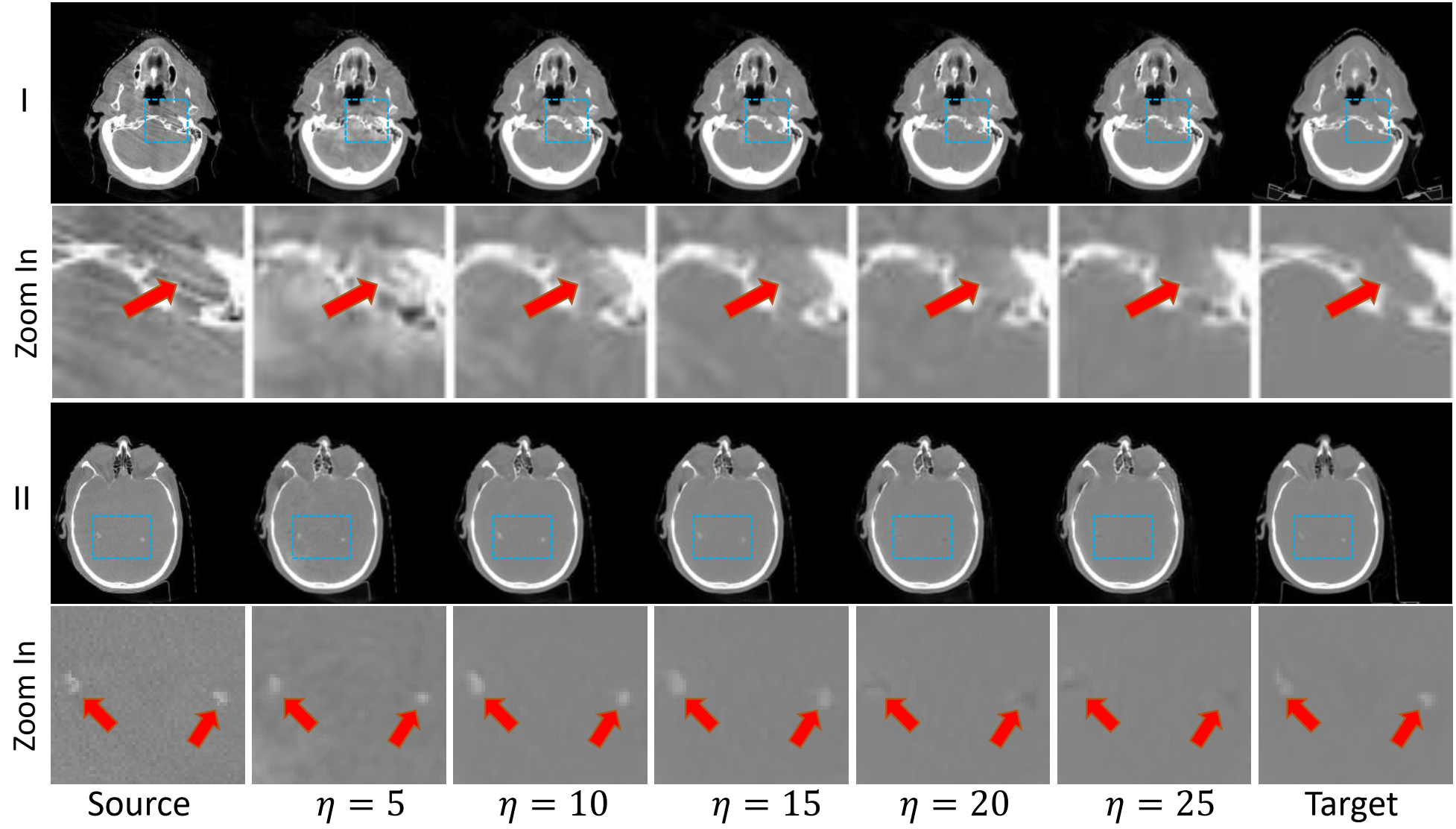}
    \caption{Visual comparison between images generated using different values of high-frequency ratio parameter $\eta$ on the head and neck dataset. The display window is [-1000,1000] HU.}
  \label{fig:eta}
\end{figure*}

\subsection{Performance on the Organs at Risk (OARs) Dataset}
The robustness and generalizability of the image translation models were further evaluated on a more challenging dataset, the Organs at Risk (OARs) dataset. The head and neck images used to train FGDM are typically surrounded by a dark background (air), whereas most of the images in the OARs dataset are not. Therefore, the significant difference in data distribution helps to test different models' robustness and generalizability.

As shown in Table \ref{tab:aapm_result} and Fig. \ref{fig:aapm_result}, the performance of most benchmark models suffered noticeably on this dataset, indicating their difficulties to generalize when challenged with a substantial shift in testing data distributions. Comparatively, the FGDM showed a substantially more robust performance across all metrics. It successfully retained the diffusion model's ability to produce high-quality and realistic images, while also preserving anatomical details through frequency-domain guidance.  Despite the significant differences between the OARs dataset and the head and neck dataset, FGDM demonstrated robustness in generalizing and adapting to substantial shifts in data distributions. \\

\subsection{Performance on the Head MR T1 Dataset}

The final evaluation of the translation models was performed on a dataset consisting of Head MR T1 images from two different centers: NFBS and ABCs. This evaluation serves the needs of image translation/harmonization, in scenarios where images from different institutions can differ in terms of imaging protocols and machine conditions. We trained translation models from NFBS to ABCs for all methods and evaluated their performance based on the FID, SSIM, and PSNR metrics. Since NFBS and ABCs are not paired data, we only tested the FID between the translated images and the target dataset to evaluate the realism, and the PSNR and SSIM between the translated images and the source images to evaluate the faithfulness. The quantitative results are shown in Table \ref{tab:mrt1} and the visual comparison in Fig. \ref{fig:mrt1_result}. FGDM consistently demonstrated superior performance on all metrics. It not only produced high-quality and realistic images but also managed to preserve the anatomical details effectively, demonstrating a robust image translation capability across the two datasets via a zero-shot framework.

\begin{table*}[ht]
\centering
\caption{Impact of the low-frequency ratio parameter $\Tilde{T}$. The arrows are pointing to the directions of improved accuracy.}
\renewcommand\arraystretch{1.1}
\setlength{\tabcolsep}{4.5mm}
\begin{tabular}{c|c|cc|cc}
\Xhline{1pt}
\multirow{2}{*}{Low Frequency Ratio (Step $\Tilde{T}$)}             & \multirow{2}{*}{FID$\downarrow$   } & \multicolumn{2}{c|}{With Source} & \multicolumn{2}{c}{With Target} \\ \cline{3-6} 
         &        & PSNR $\uparrow$ & SSIM $\uparrow$    & PSNR $\uparrow$ & SSIM $\uparrow$ \\ \hline
1 & 26.1  & \textbf{30.2} & \textbf{0.922}  & 26.4 & 0.903\\
2 &26.0 & 29.1 & 0.915  & 26.4 & 0.903 \\
3 &25.0  & 28.4 & 0.911  & 26.4 & 0.904 \\
\textbf{4} &  \textbf{24.0}  & 27.8 & 0.908 & \textbf{26.4} & \textbf{0.904} \\  
5 & 26.7  & 27.2 & 0.888 & 25.8 & 0.881\\\Xhline{1pt}
\end{tabular}
\label{tab:t}
\end{table*}

\begin{figure*}[ht]
\centering
  \includegraphics[width=.85\textwidth]{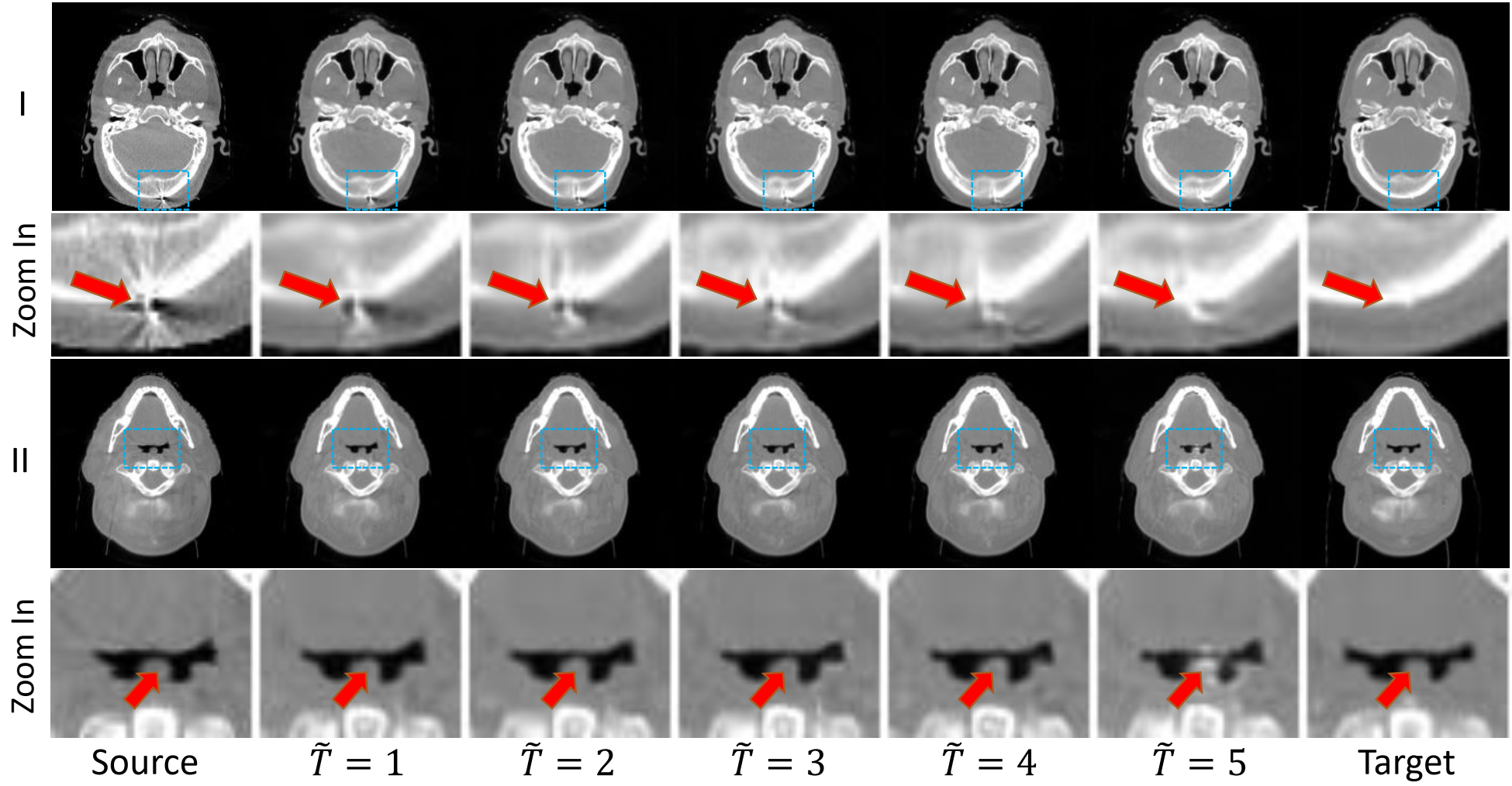}
    \caption{Visualization comparison with different low frequency ratio (step $\Tilde{T}$) on the head and neck dataset. The display window is [-1000,1000] HU.}
  \label{fig:t_study}
\end{figure*}

\subsection{Ablation Study}
\subsubsection{Effectiveness of high-frequency and low-frequency conditioning}
The proposed FGDM framework uses high-pass and low-pass filters to obtain domain-invariant information as conditioning input for the diffusion model. To test their relative contributions, we performed an ablation study with the head and neck dataset. The results of the ablation experiments are shown in Table \ref{tab:abalation}, and it can be seen that the performance of the model decreases significantly after removing the high-frequency or low-frequency information. In particular, the SSIM of the model decreases substantially after removing the high-frequency guidance.

\subsubsection{Impact of the high-frequency ratio parameter $\eta$} 
For FGDM, the high-pass filter has an adjustable $\eta$ parameter that can be customized on-the-fly during the test time to fit different image translation tasks. Since the data in the source and target domains differ in frequency space for different image translation tasks,  retaining appropriate high-frequency information can both filter out undesired information, like artifacts in CBCT, and preserve the anatomical structure. As shown in Table \ref{tab:eta}, the model achieves optimal performance when $\eta$ equals 10. As $\eta$ increases further, more information is filtered out by the high-pass filter, resulting in insufficient high-frequency guidance for structure retention, shown in Fig. \ref{fig:eta}(II). When $\eta$ is smaller than 10, the high-frequency information contains more artifacts, resulting in degraded model performance, shown in Fig. \ref{fig:eta}(I).
During the test time, $\eta$ can be freely adjusted by the user to select an optimal value for each case. 

\subsubsection{Impact of the low-frequency ratio parameter $\Tilde{T}$ } 
Similarly, FGDM can adjust the extent of retained low-frequency information by implementing various forward diffusion steps. Similar to the high-frequency counterpart, the low-frequency information can also be customized and optimized according to the image translation tasks. As shown in Table \ref{tab:t}, when the number of steps $\Tilde{T}$ is reduced, i.e., more low-frequency information is retained, the PSNR and SSIM of the translated images with the source images are higher, which is expected as the filtered image gets closer to the original CBCT domain by using fewer diffusion steps. However, the FID decreases and the PSNR and SSIM compared to the reference CT also decrease, showing the image is not well translated to match the target domain. Fig. \ref{fig:t_study} shows that when $\Tilde{T}$ is smaller, the translated image is closer to the original image, but more streak artifacts in the original image are also kept. When T equals 5, the image intensity deviates from the reference CT image. Balancing the structural similarity to the CBCT domain and the intensity similarity to the CT domain, we consider a step number $\Tilde{T}$ of 4 as optimal. 
Based on the observations, the high- and low-frequency retention factors can be properly chosen to balance the faithfulness and realism of the translated images and can be optimized on-the-fly during the inference stage with minimally incurred computational load.

\section{Limitation}
FGDM is based on the premise that the differences between the source and target domains are mainly concentrated at the intermediate frequencies, while the low-frequency information and the high-frequency information should be similar. Therefore, our model is currently only applicable for translation tasks between similar modalities such as CBCT and CT, or medical images with the same modality but different acquisition centers, machines, or protocols. For tasks such as MR-to-CT translation, where the differences between low-frequency and high-frequency information are substantial, our model may not perform well. In future work, we will develop methods to convert the low-frequency and high-frequency information between modalities like CT and MR first, before filling in the intermediate frequencies using the developed FGDM framework. In addition, we chose diffusion GAN as our backbone network as we want to achieve a balance between computational speed and image quality, as the CBCT-to-CT conversion efficiency is of critical importance for online adaptive radiotherapy. In future work, the frequency guidance using the original DDPM backbone can also be evaluated for its potential in generating more accurate images when the inference speed is not a critical factor. Another aspect is that our study only focused on the design of the overall framework and did not work on optimizing the model structure or designing more tailored loss functions for the image translation task, which can help to further improve the translation results. Future works can focus on improving loss functions to emphasize important anatomical structures for each specific translation task and optimizing the model structure to include potential attention mechanisms to enhance the accuracy and efficiency of image translation.

\section{Conclusion}

In this paper, we propose a frequency-guided diffusion model for medical image translation. Based on the observations in the frequency domain, we extract domain invariant information by high-pass and low-pass filters. The addition of high-frequency information greatly improves the retention capability of anatomical structures of the diffusion model in medical image translation tasks. Moreover, the thresholds of our two frequency filters can be freely adjusted during test-time to cope with different types of medical image translation tasks without re-training the model. FGDM not only outperforms other models in all metrics, but can also be trained only on the target-domain images and directly applied to the source-domain images to achieve zero-shot image translation. We applied the FGDM model trained on the head and neck image dataset to translate the lung image dataset and the OARs image dataset without any fine-tuning, further demonstrating the robustness of our model for zero-shot image translation and out-of-distribution data.

\bibliographystyle{IEEEtran}
\bibliography{paper.bib}

\end{document}